\documentclass[reprint]{revtex4-1}

\usepackage[pdftex]{graphicx}
\graphicspath{{./}}
\usepackage{amssymb,amsmath}

\newcommand{\Ai}{\mathop{\rm Ai}\nolimits}

\begin{document}

\title{Backscattering of Laser Radiation on Ultra-Relativistic Electrons in Transverse Magnetic Field: Evidence of Photon Interference in a MeV Scale}
\author{E. V. Abakumova}
\author{M. N. Achasov}
\author{D. E. Berkaev}
\author{V. V. Kaminsky}
\author{N.~Yu.~Muchnoi} \email{N.Yu.Muchnoi@inp.nsk.su}
\author{E. A. Perevedentsev}
\author{E. E. Pyata}
\author{Yu. M. Shatunov}
\affiliation{Budker Institute of Nuclear Physics Siberian Branch of the Russian Academy of Sciences}
\affiliation{Novosibirsk State University, 630090, Novosibirsk, Russia}

\begin{abstract}
The experiment on laser light backscattering on relativistic electrons was carried out at the VEPP-2000 collider. 
Laser radiation ($\lambda_0\simeq 10.6~\mu$m) was scattered head-on the 990~MeV electrons inside the dipole magnet, where an electron orbit radius is about 140~cm.
The energies of backscattered photons were measured by the HPGe detector.
It was observed experimentally that due to the presence of magnetic field, energy spectrum of backscattered photons differs from the Klein-Nishina cross section.
The explanation of the effect is proposed in terms of classical electrodynamics. 
Moreover, it appears that the exact QED predictions for the phenomenon were done more than 40 years ago.
\end{abstract}

\maketitle

\section{Introduction}

\begin{figure}[h]
\centering
\includegraphics[width=\linewidth]{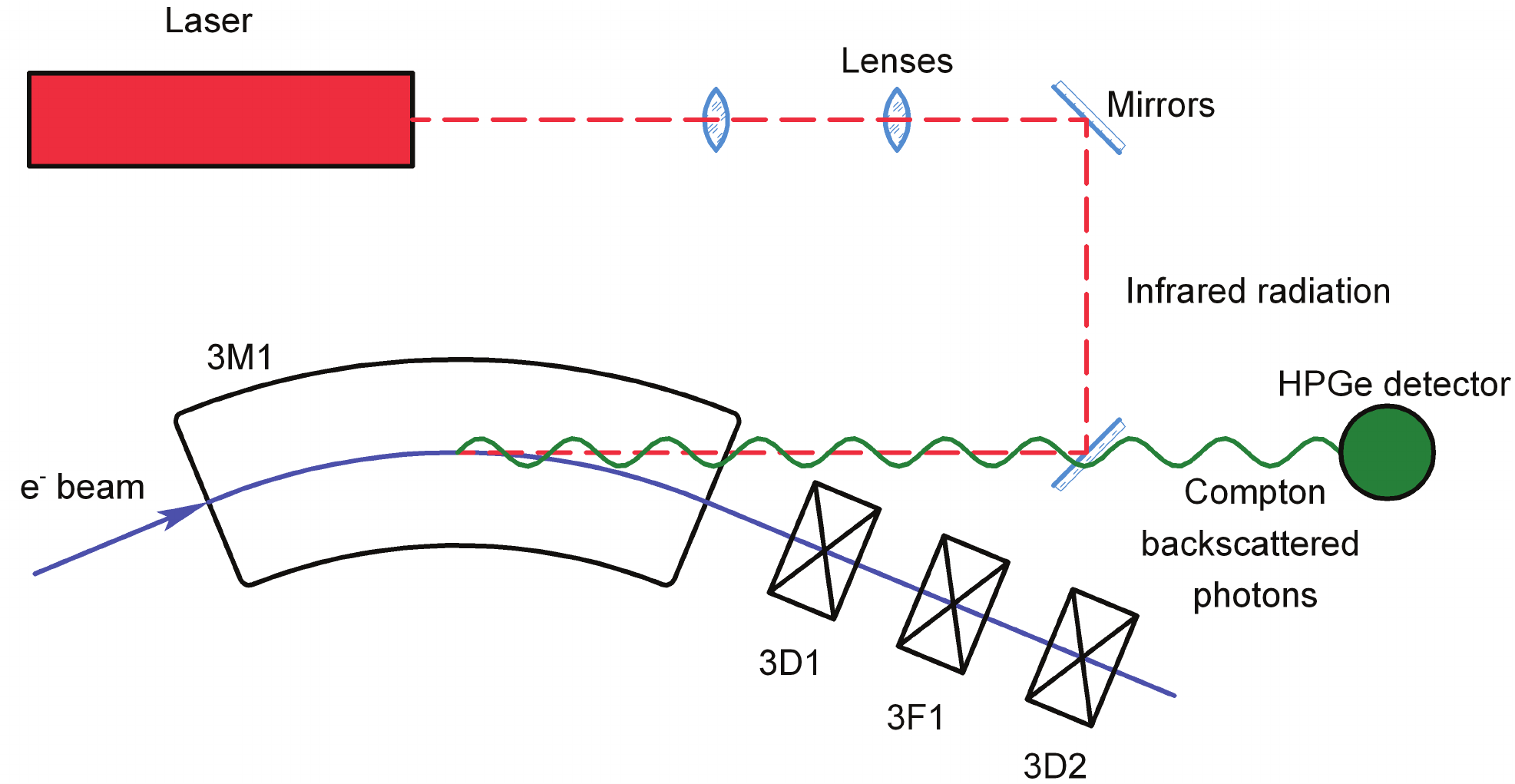}
\caption{Experimental layout. $CO_2$ Laser: Coherent GEM Select 50. HPGe detector: Ortec GMX25-70-A.}
\label{fig:experiment}
\end{figure}

VEPP-2000 \cite{V2K} is the $e^+e^-$ collider, operating in the energy range $0.2 \leq E_{CM} \leq 2.0$~GeV. 
It has 8 equal dipole magnets with electron bend radius $R=140$~cm. 
The monochromatic laser radiation ($\lambda_0=10.591035~\mu$m, CW) is injected into the collider vacuum chamber towards the electron beam according to FIG.~\ref{fig:experiment}. 
The laser and electron beams interaction occurs inside the 3M1 magnet and backscattered photons hit the HPGe detector, located in the orbit plane at the distance of $\simeq$~225~cm from the interaction area.

According to  kinematics of inverse Compton scattering, the maximum energy of backscattered photons is $\hbar\omega_{max}\simeq4\gamma^2(\hbar\omega_0)$ for head-on collision, where $\gamma = E/m$ is the electron Lorentz factor.  
In absence of constant EM field, the energy spectrum of Compton photons has an abrupt edge at this energy, which is often used as a reference energy point e.~g. for calibration purposes. 
In particular, the beam energy measurement systems, based on this principle, are widely used now in accelerator laboratories. 
So we are not going to discuss here various technical aspects like an accuracy of the HPGe detector energy scale calibration, its response function shape, etc. 
The references about these issues may be found in~\cite{BEPC} and citations therein.

\begin{figure}[h]
\centering
\includegraphics[width=\linewidth]{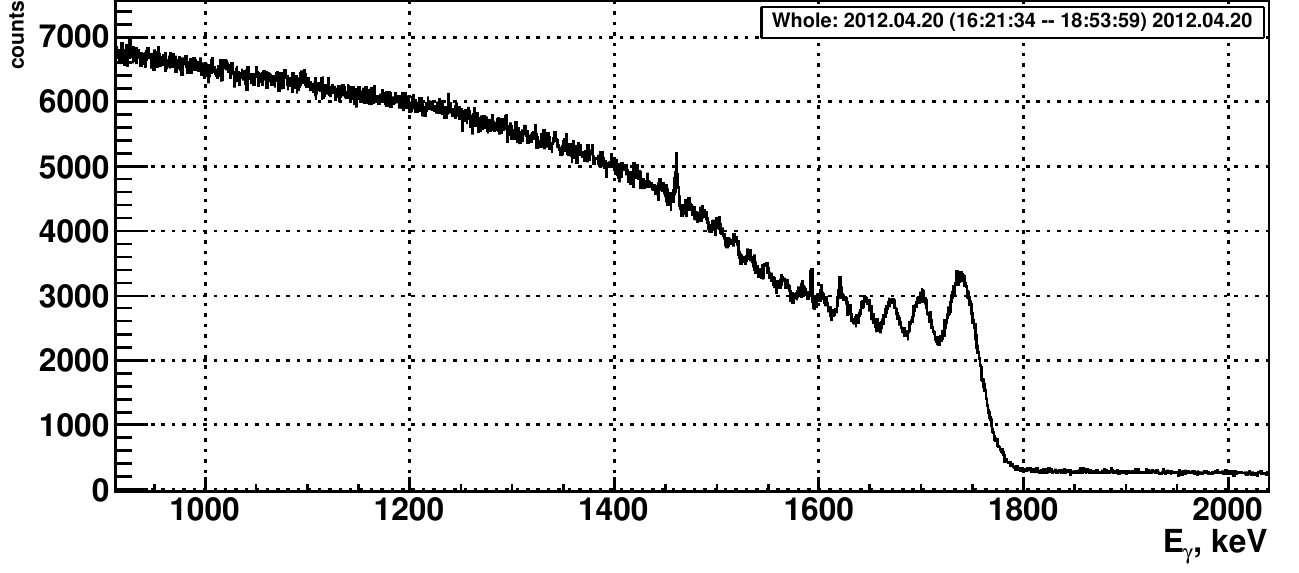}
\caption{The measured energy spectrum of backscattered photons. $E\simeq990$~MeV, $\hbar\omega_0=0.117$~eV, acquisition time 2.5~hours. Kinematics gives $\hbar\omega_{max}\simeq1755$~keV.}
\label{fig:sp-experiment}
\end{figure}

The experimental energy spectrum of laser photons, backscattered on the electron beam, is shown in FIG.~\ref{fig:sp-experiment}. 
The amplitude oscillations in the spectrum are clearly seen at least in the range from 1600~keV to 1800~keV. 

In order to understand our observations let's consider the interaction between an electron and laser wave in terms of classical electrodynamics.

\section{Interference of scattered waves}

Laser beam propagates along a tangent towards the electron beam orbit, see FIG.~\ref{fig:experiment}. 
It is focused by two ZnSe lenses providing the transverse waist size of about 1~mm. 
Thus, the approximate length of the interaction region is about $L_{int}\simeq10$~cm, that corresponds to an electron bending angle $\theta_{int} = L_{int} / R \simeq 70 $~mrad. 
Practically, the overlap of laser and electron beams is achieved by means of the positive feedback system, which finds the maximum rate of backscattered photons, counted by the HPGe detector, via fine tunning of transverse and longitudinal positions of the laser beam waist.

\begin{figure}[h]
\centering
\includegraphics[width=0.8\linewidth]{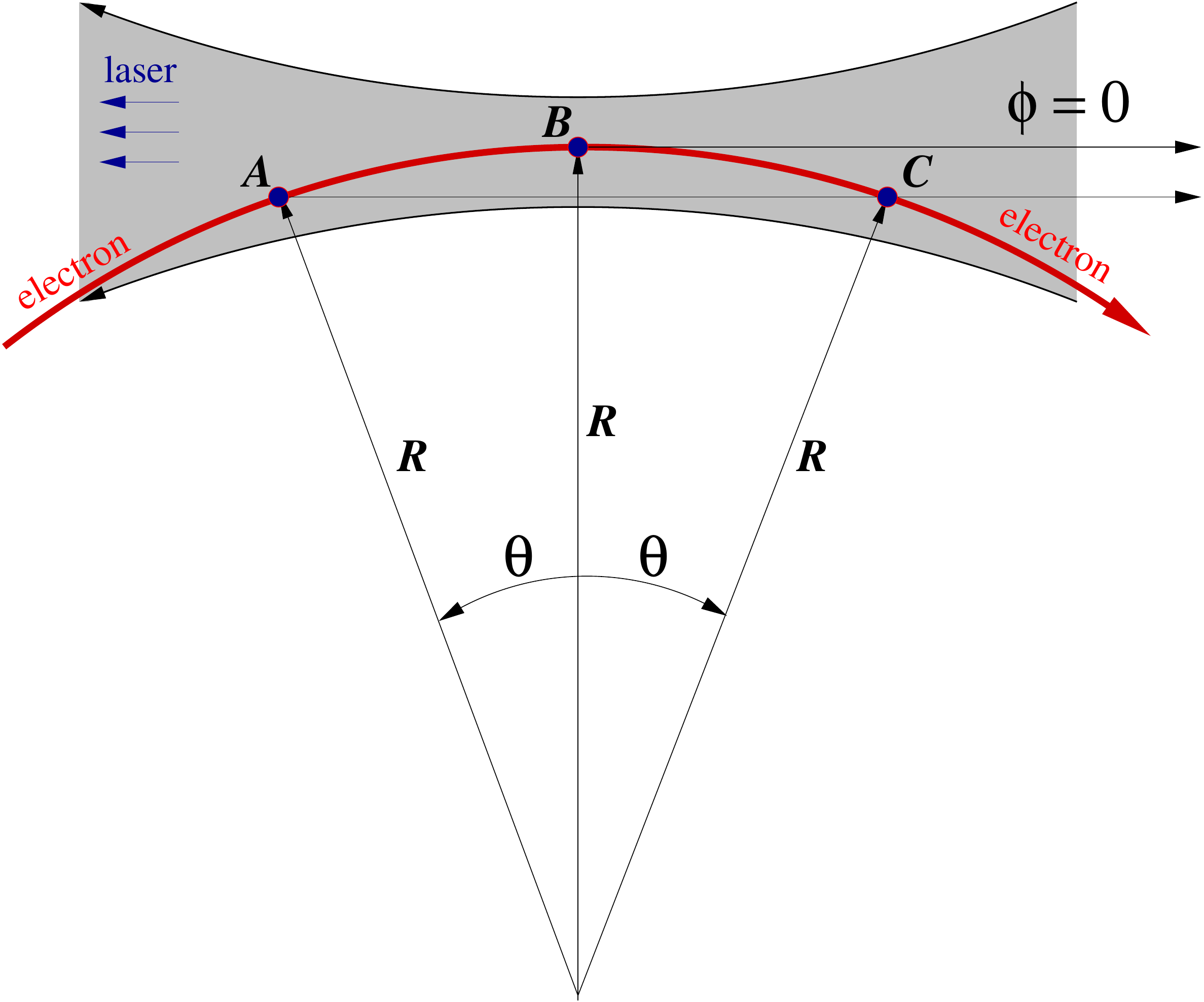}
\caption{The sketch of the laser-electron beam interaction region. $R$=140~cm is the electron orbit radius.}
\label{fig:cxema}
\end{figure}

The sketch of the interaction region is shown in FIG.~\ref{fig:cxema}. 

A typical radiation angle of an ultra-relativistic electron $\theta_{rad} \simeq 1/\gamma$, this is about 0.5~mrad if $E\simeq1$~GeV. 
Thereby $\theta_{rad}$ is two orders of magnitude less than $\theta_{int}$. 
This case is in general quite similar to the case of e.~g. synchrotron radiation in a uniform field: whatever the radiation properties are, they are the same for any direction, lying in the plane orthogonal to electron orbit plane and tangent to the electron orbit. 
In other words, it is enough to find the properties of backscattered laser radiation, propagating in the planes, orthogonal to the plane of FIG.~\ref{fig:cxema} and with $\phi=0$. 

Assuming possible interference of the waves, emitted from points $A$ and $C$ in a certain direction, it is necessary to determine the phase difference between these two waves. 
In our case the direction of radiation is specified by two angles: $\phi\equiv0$ and $\psi$, the vertical angle with respect to the orbit plane.
The time of an electron flight from point $A$ to point $C$ is:
\begin{equation}
t_e  =  \frac{2 R \theta}{v}= \frac{2 R \theta}{\beta c}.
\label{t_e}
\end{equation}
The wavefront of the radiation, emitted from point $A$ with a vertical angle $\psi$, will spatially coincide with the wavefront, emitted from point $C$ at the same angle after
\begin{equation}
t_\gamma  =  \frac{2 R \sin\theta}{c}\cos\psi.
\label{t_g}
\end{equation}
The phase difference between these waves for a certain wavelength $\lambda$ is determined by the laser wavelength $\lambda_0$:
\begin{eqnarray}
 \Delta \Phi & = & 2 \pi c \left( \frac{t_e}{\lambda} - \frac{2t_e}{\lambda_0} - \frac{t_\gamma}{\lambda} \right) = \nonumber \\
& = & \frac{2R}{c} \left( \frac{\theta}{\beta}\left(\omega-2\omega_0\right) - \omega\sin\theta\cos\psi \right),
\label{phase}
\end{eqnarray}
where we take into account the laser wave phase shift while the electron propagates from $A$ to $C$.
Since $\theta, \psi, 1/\gamma \ll 1$, Eq.~(\ref{phase}) transforms to:
\begin{equation}
\Delta \Phi \simeq \frac{\omega R}{c} \left( 
\theta\left(\frac{1}{\gamma^2} - \frac{4\omega_0}{\omega} + \psi^2\right) + \frac{\theta^3}{3} \right).
\label{phase_final}
\end{equation}
When $\Delta\Phi$ is an odd (even) multiple of $\pi$ one observes an interference minimum (maximum) of the scattered wave. 
The scattered field amplitude is evaluated by integration along the electron path:
\begin{equation}
U  \propto  \omega \int \limits_{0}^{\infty} \left( e^{i\frac{\Delta\Phi}{2}}+e^{-i\frac{\Delta\Phi}{2}} \right) d\theta = 
2 \omega \int \limits_{0}^{\infty}\cos{\frac{\Delta \Phi}{2}} d\theta.
\label{Guigens}
\end{equation}
Change of the integration variable $\theta \rightarrow \xi = \theta \bigl( \omega R/2 c \bigr)^{1/3}$ allows to rewrite expression (\ref{Guigens}) as:
\begin{equation}
U \propto \omega^{2/3} \Ai(x),
\label{Airy}
\end{equation}
where
\begin{equation}
x = \left(\frac{ \omega R}{2c}\right)^{2/3}\left(\frac{1}{\gamma^2} - \frac{4\omega_0}{\omega} + \psi^2\right),
\label{X}
\end{equation}
and $\Ai(x) = \displaystyle\frac{1}{\pi} \int \limits_{0}^{\infty}\cos{(xt+\frac{t^3}{3})} dt$ is the Airy function.
The intensity of scattered wave is:
\begin{equation}
I = |U|^2 \propto \omega^{4/3} \Ai^2(x).
\label{Airy2}
\end{equation}

Expression (\ref{Airy2}) is the solution for the angular distribution of the backscattered radiation spectral power density. 
According to the above approach, the result does not depend on a particle type the laser wave is scattered on. 
Similar solution may be obtained by the analysis of the Fourier harmonics of the radiation field of charged particle~\cite{landau1975classical}.

To obtain the energy spectrum of scattered photons one should integrate expression (\ref{Airy2}) over the vertical angle $\psi$ and divide the result by the photon energy $\hbar\omega$:
\begin{equation}
\frac{d\dot{N}_\gamma}{d\hbar\omega} \propto \omega^{1/3} \int \limits_{0}^{\infty} \Ai^2(x) d\psi.
\label{cspectrum0}
\end{equation}
Integral in Eq.~(\ref{cspectrum0}) can be expressed via the primitive of  Airy function using the relation:
\begin{equation}
\int \limits_{0}^{\infty} \Ai^2(a+by^2) dy =
\frac{1}{4\sqrt{b}}\int \limits_{z}^{\infty} \Ai(z') dz',\;\;z=2^{2/3}a.
\label{Airy-transform}
\end{equation}
Hence, the final form of the interference factor is:
\begin{equation}
\frac{d\dot{N}_\gamma}{d\hbar\omega} \propto \int \limits_{z}^{\infty} \Ai(z') dz'
= \frac{1}{3}-\int \limits_{0}^{z} \Ai(z') dz',
\label{cspectrum1}
\end{equation}
where
\begin{equation}
z = \Bigl(\frac{ \omega R}{c}\Bigr)^{2/3}\Bigl(\frac{1}{\gamma^2} - \frac{4\omega_0}{\omega}\Bigr).
\label{Z}
\end{equation}

\begin{figure}[ht]
\centering
\includegraphics[width=1.02\linewidth]{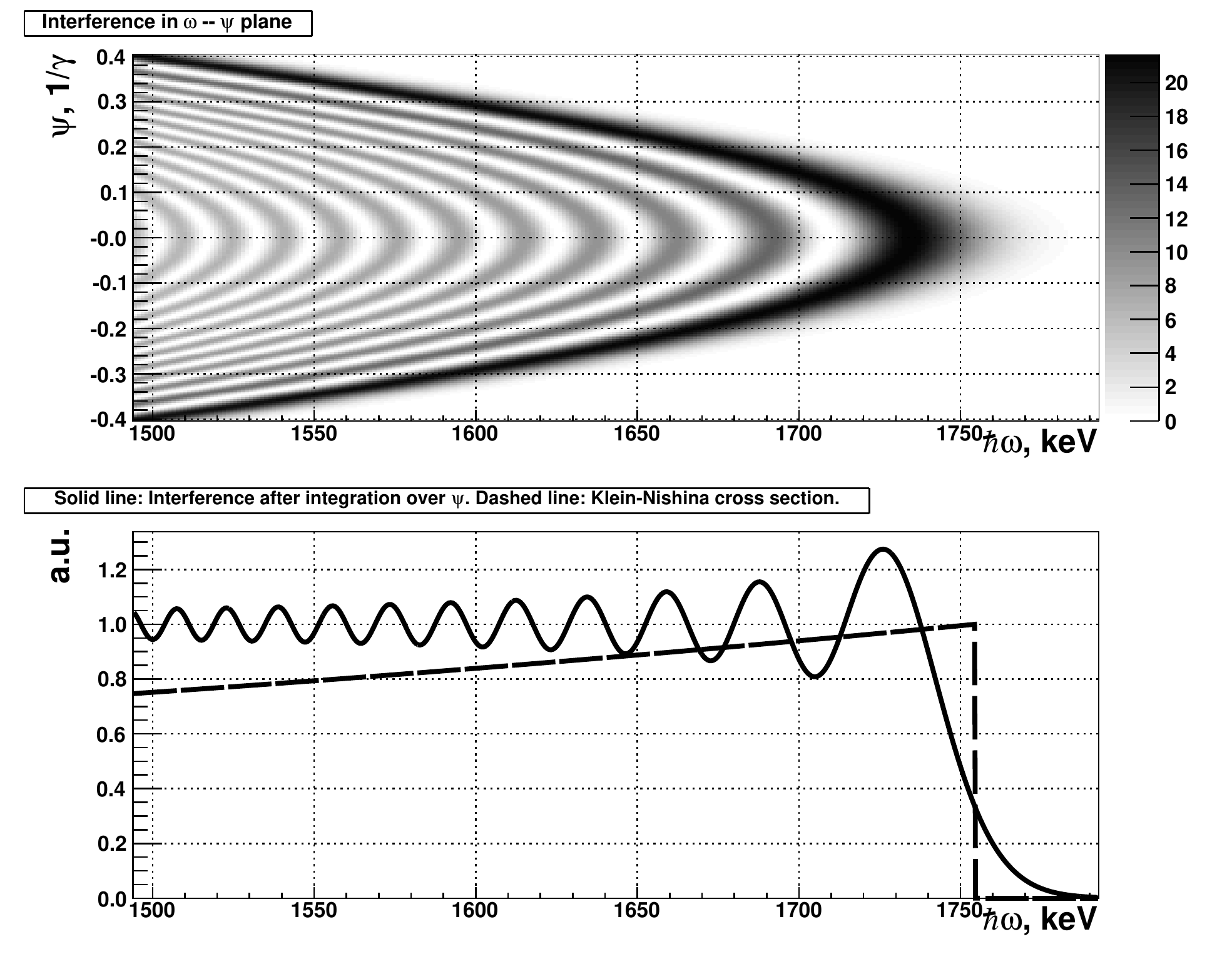}
\caption{Upper plot: interference of scattered waves according to Eq.~(\ref{Airy2}) in $\omega - \psi$ plane. 
Lower plot: solid line -- the energy distribution of backscattered photons according to Eq.~(\ref{cspectrum1}), dashed line -- the same according to Klein-Nishina cross section with the abrupt edge given by the Compton scattering kinematics. 
$E=990$~MeV, $\hbar\omega_0=0.117$~eV, $R=140$~cm.}
\label{fig:2D}
\end{figure}

The results, represented by Eq.~(\ref{Airy2}) and Eq.~(\ref{cspectrum1}), are shown in FIG.~\ref{fig:2D}. 
The 2D distribution of scattered wave intensity in $\omega-\psi$ plane shows the sought-for interference effect with 100~\% intensity modulation. 
After integration over $\psi$, the interference is still evident in the energy spectrum of scattered photons.

Let's point out, that in presence of constant EM field in the scattering area, the abrupt high-energy edge in the energy spectrum of backscattered photons does not exist any more, see FIG.~\ref{fig:2D}.

Till now we have not yet taken into account the quantum recoil. To do this one should, according to~\cite{REL1973}, substitute $\omega \rightarrow \omega \cdot E/(E-\hbar\omega)$ in Eq.~(\ref{Z}).

An electron radius is coupled with its energy and magnetic field strength by the balance between Lorenz and centrifugal forces: $\beta E = c B R$. 
It is convenient to perform $ R \rightarrow E / (cB) $ substitution in Eq.~(\ref{Z}).

After these substitutions were made, let's introduce new variables:
\begin{equation}
u      = \frac{\hbar\omega}{E-\hbar\omega},\;\;\;
\kappa = \frac{4E\hbar\omega_0}{m^2},\;\;\;
\chi   = \frac{E}{m} \frac{B}{B_0},
\label{ukh}
\end{equation}
where $B_0 = m^2/\hbar c^2 = 4.414\cdot10^{9}$~T is the Schwinger field strength. 
Now expression (\ref{Z}) looks like:
\begin{equation}
z = (u/\chi)^{2/3} (1 - \kappa/u)
\label{Znew}
\end{equation}

The spectrum shape, similar to Eq.~(\ref{cspectrum1}), was obtained in semiclassical theory of electromagnetic processes~\cite{BKS-1991}.

\section{QED cross section}
The influence of constant EM field on the cross section of Compton scattering was studied in~\cite{Zhukovsky-Hermann}, where
the energy spectrum of scattered photons was obtained from the solution of the Dirac equation: 
\begin{equation}
\frac{d\dot{N}_\gamma}{d\hbar\omega} \propto \nu_1 \int\limits^{\infty}_{z} \Ai(z') dz' + \nu_2 \Ai'(z) + \nu_3 \Ai(z),
\label{spectrum_q}
\end{equation}
where
\begin{eqnarray}
\nu_1 &=& \frac 1 8 \left\{ 2 + \frac{u^2}{1+u} - 4 \frac u \kappa + 4\left[ \frac u \kappa \right]^2 - 16 \left[ \frac u \kappa \right]^2 \left[ \frac \chi \kappa \right]^2 \right\}, \nonumber\\
\nu_2 &=& -\left[ \frac u \kappa \right]^{\frac{4}{3}} \left[ \frac \chi \kappa \right]^{\frac{2}{3}} 
\left\{ 4 \left[ \frac \chi \kappa \right]^2 + \frac{u^2}{2(1+u)} \left[ 1 + 4 \left[ \frac \chi \kappa \right]^2 \right] \right\}, \nonumber\\
\nu_3 &=& \left[ \frac u \kappa \right]^{\frac{2}{3}} \left[ \frac \chi \kappa \right]^{\frac{4}{3}} 
\left\{ 3 - 2 \frac u \kappa + \frac{u^2}{2(1+u)} \left[3 - 4 \frac u \kappa \right] \right\}.
\label{nu_123}
\end{eqnarray}

In case relevant to our experiment $u \lesssim 10^{-3}$, $\kappa \lesssim 2 \cdot 10^{-3}$ and $\chi \lesssim 10^{-6}$.
The influence of constant field on the process of Compton scattering depends on $\chi/\kappa$ ratio.
Since $u/\kappa \simeq 0.5$ and $\chi/\kappa \lesssim 0.5\cdot10^{-3}$, the last term in $\nu_1$ in Eq.~(\ref{nu_123}) may by omitted, and $\nu_1$ becomes just the Klein-Nishina cross section. 
The values of $\nu_2$ and $\nu_3$ in Eq.~(\ref{nu_123}) are significantly smaller than $\nu_1$ due to the same reason. 
In this case one can see that the QED result is the product of Eq.~(\ref{cspectrum1}) by the Klein-Nishina cross section.

\section{Results and Discussion}

Now we are going to compare the measured energy spectrum of backscattered photons with the theory predictions. 
In order to do this it is necessary to take into account the energy spread of the electrons in the beam. 

Let $\delta$ be the relative energy shift of an electron energy $E'$ from the average energy $E$ of electrons in the beam: $E'=E(1+\delta)$. 
The appropriate weight function for the electrons energy distribution would be:
\begin{equation}
w(\delta) = \frac{1}{\sqrt{2\pi}\sigma}\exp{\left(-\frac{\delta^2}{2\sigma^2}\right)},
\label{spread}
\end{equation}
where $\sigma$ is by definition the relative beam energy spread. 
In case $\sigma \ll 1$ the linear approximation for the coupling between $z$ and $\delta$ would be adequate. From Eq.~(\ref{Znew}):
\begin{equation}
z(\delta) \simeq z - \eta \cdot \frac \delta \sigma,\;\;\eta \simeq \sigma \cdot \frac 4 3 \left(1 + \frac 1 2 \frac \kappa u \right) \left( \frac u \chi \right)^{2/3}.
\label{zspread}
\end{equation}
The energy spectrum transformation due to non-zero energy spread in the electron beam is established by convolution of Eq.~(\ref{cspectrum1}) and Eq.~(\ref{spread}), yielding the final result:
\begin{equation}
\frac{d\dot{N}_\gamma}{d\hbar\omega} \propto 
\mathcal{F}(\omega,E,B,\sigma) =
e^{- \frac{\eta^6}{24}}
\int \limits_{z+ \eta^4/4}^{\infty} e^{\frac{z' \eta^2}{2}}\Ai(z') dz'.
\label{sspread}
\end{equation}

Let's introduce a combined function to describe the shape of the experimental spectrum:
\begin{equation}
f(\omega) = \mathcal{A}\cdot\mathcal{F}(\omega,E,B,\sigma) + \mathcal{B}(\omega),
\label{fitf}
\end{equation}
where $\mathcal{B}(\omega) = p_0 + p_1(\omega-\omega_{max})$ is the estimation of the background shape.
$f(\omega)$ has 6 free parameters:
\begin{itemize}
\item $\mathcal{A}$ is the spectrum normalization parameter,
\item $E$ is the average energy of electrons in the beam,
\item $B$ is the magnetic field in the interaction area,
\item $\sigma$ is the relative electron energy spread,
\item $p_0$ and $p_1$ describe the linear background,
\end{itemize}
while the laser photon energy $\hbar\omega_0=0.117065223$~eV.
\begin{figure}[h]
\centering
\includegraphics[width=\linewidth]{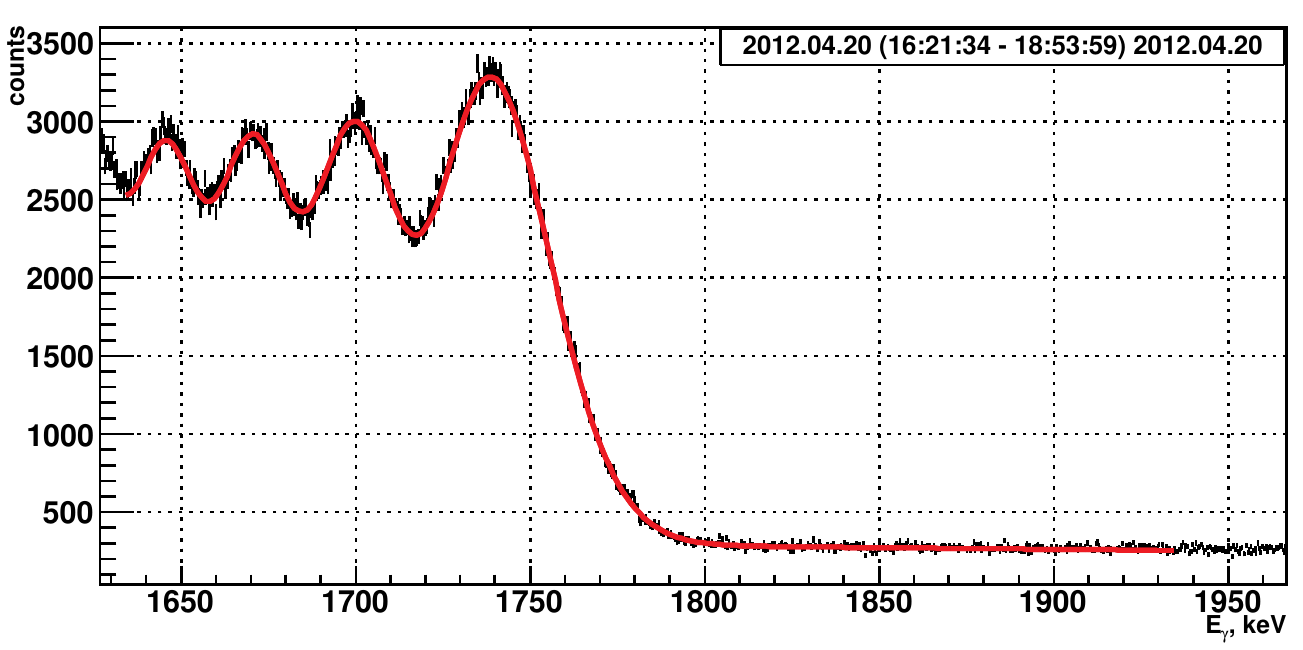}
\caption{The energy spectrum with the fit function. }
\label{fig:edge-with-fit}
\end{figure}
\begin{table}[h]
\centering
\begin{tabular}{|l|l c l|}
\hline
~Parameter name~   & \multicolumn{3}{|c|}{Parameter value}    \\ \hline
~$\mathcal{A}$      & ~2378.87 &$\pm$& ~4.64    ~\\
~$E$, MeV           & ~993.662 &$\pm$& ~0.016   ~\\
~$B$, T             & ~2.38802 &$\pm$& ~0.00442 ~\\
~$\sigma$           & ~0.00081 &$\pm$& ~0.00004 ~\\
~$p_0$              & ~295.94  &$\pm$& ~2.10    ~\\
~$p_1$, keV$^{-1}$  & ~-0.212  &$\pm$& ~0.022   ~\\ \hline
\end{tabular}
\caption{Fit results: $\chi^2/NDF=773.0/745$, Prob. = 0.231.}
\label{tab:table}
\end{table}

The edge of the experimental spectrum is well fitted with $f(\omega)$, see FIG.~\ref{fig:edge-with-fit} and TABLE~\ref{tab:table}.
If we use  Eq.~(\ref{spectrum_q}) instead of Eq.~(\ref{cspectrum1}) for experimental data approximation, i.~e. take into account Klein-Nishina formula, then the fit has a low confidence level $<0.001$. 
The overall shape of the measured energy spectrum (see FIG.~\ref{fig:sp-experiment}) is rather complex and depends on many experimental circumstances such as the spectra of backscattered photons and  bremsstrahlung radiation, $\gamma$-lines from nuclear reactions in the experimental area, Compton scattering of the $\gamma$-quanta inside the detector as well as in the medium between the interaction area and the detector, etc. However, in the narrow energy range near the highest energies of the scattering photons, the distortion of the spectrum shape due to these effects is relatively low.

The fitting results in TABLE~\ref{tab:table} indicate that the properties of the observed phenomenon provide an opportunity for simultaneous measurement of three parameters: 
\begin{itemize}
\item average electron energy $E$, with relative statistical accuracy $\Delta E / E \simeq 2 \cdot 10^{-5}$,
\item relative energy spread in the electron beam $\sigma$, $\Delta\sigma / \sigma \simeq 4\%$,
\item magnetic filed strength in the interaction area $B$,  $\Delta B / B \simeq 0.2\%$.
\end{itemize}
In order to implement this further careful studies of the spectrum shape are required.

\section{Conclusion}

The laser backscattering experiment, held on the VEPP-2000 collider, demonstrates the influence of the constant EM field on the interaction of photons with relativistic electrons.
The experimental results are consistent with either QED or classical theory predictions: the observed phenomenon can be explained as the interference of photons with $\lambda \sim 10^{-10}$~cm. 
The peculiar scattering geometry looks promising as a new diagnostic tool for accelerator-based experiments.

The authors are grateful to V.~E.~Blinov, A.~A.~Korol, A.~A.~Krasnov, A.~G.~Kharlamov, S.~I.~Serednyakov, Yu.~A.~Tikhonov and V.~M.~Tsukanov for support of this work, to Z.~K.~Silagadze, A.~I.~Milstein and V.~M.~Strakhovenko for useful discussions. The work is supported by the Ministry of Education and Science of the Russian Federation, by RF Presidential Grant for Scientific Schools NSh-6943.2010.2, by RFBR grants 12-02-01250-a, 12-02-00065-a, 11-02-00276-a.

\bibliography{ems}

\begin{thebibliography}{1}%
\makeatletter
\providecommand \@ifxundefined [1]{%
 \ifx #1\undefined \expandafter \@firstoftwo
 \else \expandafter \@secondoftwo
\fi
}%
\providecommand \@ifnum [1]{%
 \ifnum #1\expandafter \@firstoftwo
 \else \expandafter \@secondoftwo
\fi
}%
\providecommand \enquote [1]{``#1''}%
\providecommand \bibnamefont  [1]{#1}%
\providecommand \bibfnamefont [1]{#1}%
\providecommand \citenamefont [1]{#1}%
\providecommand\href[0]{\@sanitize\@href}%
\providecommand\@href[1]{\endgroup\@@startlink{#1}\endgroup\@@href}%
\providecommand\@@href[1]{#1\@@endlink}%
\providecommand \@sanitize [0]{\begingroup\catcode`\&12\catcode`\#12\relax}%
\@ifxundefined \pdfoutput {\@firstoftwo}{%
 \@ifnum{\z@=\pdfoutput}{\@firstoftwo}{\@secondoftwo}%
}{%
 \providecommand\@@startlink[1]{\leavevmode\special{html:<a href="#1">}}%
 \providecommand\@@endlink[0]{\special{html:</a>}}%
}{%
 \providecommand\@@startlink[1]{%
  \leavevmode
  \pdfstartlink
   attr{/Border[0 0 1 ]/H/I/C[0 1 1]}%
   user{/Subtype/Link/A<</Type/Action/S/URI/URI(#1)>>}%
  \relax
 }%
 \providecommand\@@endlink[0]{\pdfendlink}%
}%
\providecommand \url  [0]{\begingroup\@sanitize \@url }%
\providecommand \@url [1]{\endgroup\@href {#1}{\urlprefix}}%
\providecommand \urlprefix [0]{URL }%
\providecommand \Eprint[0]{\href }%
\@ifxundefined \urlstyle {%
  \providecommand \doi [1]{doi:\discretionary{}{}{}#1}%
}{%
  \providecommand \doi [0]{doi:\discretionary{}{}{}\begingroup
  \urlstyle{rm}\Url }%
}%
\providecommand \doibase [0]{http://dx.doi.org/}%
\providecommand \Doi[1]{\href{\doibase#1}}%
\providecommand \bibAnnote [3]{%
  \BibitemShut{#1}%
  \begin{quotation}\noindent
    \textsc{Key:}\ #2\\\textsc{Annotation:}\ #3%
  \end{quotation}%
}%
\providecommand \bibAnnoteFile [2]{%
  \IfFileExists{#2}{\bibAnnote {#1} {#2} {\input{#2}}}{}%
}%
\providecommand \typeout [0]{\immediate \write \m@ne }%
\providecommand \selectlanguage [0]{\@gobble}%
\providecommand \bibinfo [0]{\@secondoftwo}%
\providecommand \bibfield [0]{\@secondoftwo}%
\providecommand \translation [1]{[#1]}%
\providecommand \BibitemOpen[0]{}%
\providecommand \bibitemStop [0]{}%
\providecommand \bibitemNoStop [0]{.\EOS\space}%
\providecommand \EOS [0]{\spacefactor3000\relax}%
\providecommand \BibitemShut [1]{\csname bibitem#1\endcsname}%
\bibitem{V2K}%
  \BibitemOpen
  \bibfield{author}{%
  \bibinfo {author} {\bibfnamefont{D.~E.}\ \bibnamefont{Berkaev}}, \bibinfo
  {author} {\bibfnamefont{D.~B.}\ \bibnamefont{Shwartz}}, \bibinfo {author}
  {\bibfnamefont{P.~Y.}\ \bibnamefont{Shatunov}}, \bibinfo {author}
  {\bibfnamefont{Y.~A.}\ \bibnamefont{Rogovskii}}, \bibinfo {author}
  {\bibfnamefont{A.~L.}\ \bibnamefont{Romanov}}, \emph{et~al.},\ }%
  \bibfield{journal}{%
  \bibinfo {journal} {JETP}\ }%
  \textbf{\bibinfo {volume} {113}},\ \bibinfo {pages} {213} (\bibinfo {year}
  {2011})%
  \bibAnnoteFile{NoStop}{V2K}%
\bibitem{BEPC}%
  \BibitemOpen
  \bibfield{author}{%
  \bibinfo {author} {\bibfnamefont{E.~V.}\ \bibnamefont{Abakumova}}, \bibinfo
  {author} {\bibfnamefont{M.~N.}\ \bibnamefont{Achasov}}, \bibinfo {author}
  {\bibfnamefont{V.~E.}\ \bibnamefont{Blinov}}, \bibinfo {author}
  {\bibfnamefont{X.}~\bibnamefont{Cai}}, \bibinfo {author}
  {\bibfnamefont{H.~Y.}\ \bibnamefont{Dong}}, \emph{et~al.},\ }%
  \bibfield{journal}{%
  \bibinfo {journal} {Nucl. Instr. Meth.}\ }%
  \textbf{\bibinfo {volume} {A 659}},\ \bibinfo {pages} {21} (\bibinfo {year}
  {2011})%
  \bibAnnoteFile{NoStop}{BEPC}%
\bibitem{landau1975classical}%
  \BibitemOpen
  \bibfield{author}{%
  \bibinfo {author} {\bibfnamefont{L.~D.}\ \bibnamefont{Landau}}\ and\ \bibinfo
  {author} {\bibfnamefont{E.~M.}\ \bibnamefont{Lifshit͡s}},\ }%
  \emph{\bibinfo {title} {The Classical Theory of Fields}},\ \bibinfo {series}
  {Course of Theoretical Physics}, Vol.~\bibinfo {volume} {2}\ (\bibinfo
  {publisher} {Butterworth-Heinemann},\ \bibinfo {year} {1975})\ pp.\ \bibinfo
  {pages} {176--187}%
  \bibAnnoteFile{NoStop}{landau1975classical}%
\bibitem{REL1973}%
  \BibitemOpen
  \bibfield{author}{%
  \bibinfo {author} {\bibfnamefont{V.~N.}\ \bibnamefont{Baier}}, \bibinfo
  {author} {\bibfnamefont{V.~M.}\ \bibnamefont{Katkov}},\ and\ \bibinfo
  {author} {\bibfnamefont{V.~S.}\ \bibnamefont{Fadin}},\ }%
  \emph{\bibinfo {title} {Radiation by Relativistic Electrons}}\ (\bibinfo
  {publisher} {Atomizdat},\ \bibinfo {address} {Moscow},\ \bibinfo {year}
  {1973})%
  \bibAnnoteFile{NoStop}{REL1973}%
\bibitem{BKS-1991}%
  \BibitemOpen
  \bibfield{author}{%
  \bibinfo {author} {\bibfnamefont{V.~N.}\ \bibnamefont{Baier}}, \bibinfo
  {author} {\bibfnamefont{V.~M.}\ \bibnamefont{Katkov}},\ and\ \bibinfo
  {author} {\bibfnamefont{V.~M.}\ \bibnamefont{Strakhovenko}},\ }%
  \bibfield{journal}{%
  \bibinfo {journal} {JETP}\ }%
  \textbf{\bibinfo {volume} {73}},\ \bibinfo {pages} {945} (\bibinfo {year}
  {1991})%
  \bibAnnoteFile{NoStop}{BKS-1991}%
\bibitem{Zhukovsky-Hermann}%
  \BibitemOpen
  \bibfield{author}{%
  \bibinfo {author} {\bibfnamefont{V.~C.}\ \bibnamefont{Zhukovsky}}\ and\
  \bibinfo {author} {\bibfnamefont{I.}~\bibnamefont{Herrmann}},\ }%
  \bibfield{journal}{%
  \bibinfo {journal} {Journal of Nuclear Physics}\ }%
  \textbf{\bibinfo {volume} {14}},\ \bibinfo {pages} {150} (\bibinfo {year}
  {1971})%
  \bibAnnoteFile{NoStop}{Zhukovsky-Hermann}%
\end{thebibliography}%

\end{document}